# TIME DEPENDENT MODELS OF FLARES FROM SAGITTARIUS A*

Katie Dodds-Eden[1], Prateek Sharma[2,3], Eliot Quataert[3,4], Reinhard Genzel[1,4], Stefan Gillessen[1], Frank Eisenhauer[1], Delphine Porquet[5]

## ABSTRACT

The emission from Sgr A*, the supermassive black hole in the Galactic Center, shows order of magnitude variability ("flares") a few times a day that is particularly prominent in the near-infrared (NIR) and X-rays. We present a time-dependent model for these flares motivated by the hypothesis that dissipation of magnetic energy powers the flares. We show that episodic magnetic reconnection can occur near the last stable circular orbit in time-dependent magnetohydrodynamic simulations of black hole accretion – the timescales and energetics of these events are broadly consistent with the flares from Sgr A*. Motivated by these results, we present a spatially one-zone time-dependent model for the electron distribution function in flares, including energy loss due to synchrotron cooling and adiabatic expansion. Synchrotron emission from transiently accelerated particles can explain the NIR/X-ray lightcurves and spectra of a luminous flare observed 4 April 2007. A significant decrease in the magnetic field strength during the flare (coincident with the electron acceleration) is required to explain the simultaneity and symmetry of the simultaneous lightcurves. Our models predict that the NIR and X-ray spectral indices are related by $\Delta\alpha \simeq 0.5$ (where $\nu F_\nu \propto \nu^\alpha$) and that there is only modest variation in the spectral index during flares. We also explore implications of this model for longer wavelength (radio-submm) emission seemingly associated with X-ray and NIR flares; we argue that a few hour decrease in the submm emission is a more generic consequence of large-scale magnetic reconnection than delayed radio emission from adiabatic expansion.

*Subject headings:* accretion, accretion disks — black hole physics — infrared: general — radiation mechanisms: general — Galaxy: center — X-rays: general

## 1. INTRODUCTION

The monitoring of stellar orbits has established beyond reasonable doubt that the Galactic Center hosts a supermassive black hole whose mass is $\approx 4 \times 10^6 M_\odot$ (Schödel et al. 2002; Ghez et al. 2003). Observations from the radio to the X-rays reveal that coincident with the black hole is a weak active galactic nucleus (Sgr A*) whose broadband non-thermal spectrum peaks in the sub-mm at $\sim 10^{12}$ Hz (Zylka, Mezger, & Lesch 1992). The total luminosity of Sgr A* ($\sim 300 L_\odot$) is five orders of magnitude smaller than would be produced by accretion of ambient gas at the Bondi rate with a radiative efficiency of $\sim 10\%$ (e.g., Baganoff et al. 2003). Many nearby galaxies host supermassive black holes that are comparably underluminous (Di Matteo, Carilli, & Fabian 2001; Ho 2008). Thus, Sgr A* has become a critical testing ground for theoretical models of hot, radiatively inefficient accretion flows (RIAFs) that are common in the local universe.

In addition to a baseline level of quiescent emission, Sgr A* also shows short-timescale "flares" in the X-ray (Baganoff et al. 2001), near-infrared (NIR; Genzel et al. 2003), and sub-mm (Zhao et al. 2003). The duration

Electronic address: katie@mpe.mpg.de
[1] Max Planck Institut für Extraterrestrische Physik, Postfach 1312, D-85741, Garching, Germany
[2] Chandra Fellow
[3] Astronomy Department and Theoretical Astrophysics Center, University of California, Berkeley, CA 94720
[4] Department of Physics, University of California, Berkeley, 366 Le Conte Hall, Berkeley, CA 94720-7300
[5] Observatoire astronomique de Strasbourg, Université de Strasbourg, CNRS, INSU, 11 rue de l'Université, F-67000 Strasbourg, France

of the NIR ($\sim 80$ min) and X-ray flares ($\sim 50$ min) is comparable to the orbital period of matter near the last stable circular orbit around the black hole. The flare properties can thus help constrain the physical processes occurring close to the event-horizon of Sgr A*. There is currently some debate as to whether the high amplitude, short time-scale variability from Sgr A* truly consists of distinct "flares," or is instead the tail end of a power spectrum of variability (Meyer et al. 2008); for our purposes, these distinctions are not that critical and we shall refer to the high amplitude tail of Sgr A*'s variability as "flaring."

Flares from Sgr A* have been observed for about ten years and it is now possible to summarize properties common to most of them (for a more detailed discussion see, e.g., Dodds-Eden et al. 2009). Flares are more common in the NIR than in the X-ray. The flux can increase by up to a factor of $\sim 20$ above the detection limit in the NIR and up to a factor of $\sim 160$ in the X-rays. When both are present, the flares in the X-ray and NIR do not show a significant time-lag. The NIR flares are polarized, consistent with a synchrotron origin; the polarization angle can even change significantly during the flare (Eckart et al. 2006; Trippe et al. 2007; Meyer et al. 2006). For relatively luminous flares (which have better statistics) the spectrum in the NIR is $\nu L_\nu \propto \nu^{0.4}$ (Gillessen et al. 2006; Hornstein et al. 2007; Dodds-Eden et al. 2009). The bright X-ray flares show a (well constrained) softer spectrum, with $\nu L_\nu \propto \nu^{-0.25}$ (Porquet et al. 2003, 2008). The fainter X-ray flares may be harder (Baganoff et al. 2001). Better statistics are however required to conclude if there are indeed two populations of X-ray flares. (Porquet et al. 2008). Flares observed in the sub-mm ten-tatively show a lag of ∼ 100 minutes with respect to the NIR and X-ray flares (Marrone et al. 2008; Yusef-Zadeh et al. 2008).

The aim of this paper is to develop a time-dependent model of the emission from Sgr A*. Although there have been a number of models presented for the average properties of the flares (e.g., Markoff et al. 2001; Yuan et al. 2004; Liu et al. 2006), only recently has there been work studying the time-dependent emission in detail (Chan et al. 2009; Dexter, Agol, & Fragile 2009; Maitra et al. 2009). In this paper, we model the time-dependent emission from Sgr A* using a simplified model for the evolution of the electron distribution during a flare, which takes into account synchrotron cooling and other processes (e.g., adiabatic losses and escape). Our methodology complements more detailed treatments of the time-dependent emission from accretion disk simulations (e.g., Dexter, Agol, & Fragile 2009), which focus on the thermal plasma; by contrast, we model the full electron distribution function, at the expense of considering a one-zone model with no dynamics and with a specified magnetic field and size. Given the electron distribution function, we then calculate the resulting time-dependent radio to X-ray spectrum.

The observation of a rising $\nu L_\nu$ spectrum in the NIR requires that the peak synchrotron frequency for the emitting particles be $\gtrsim 10^{14}$ Hz, which in turn requires particles with Lorentz factors $\gamma \gtrsim 1000(B/30\,\mathrm{G})^{-1/2}$ ($B$ is the magnetic field strength in Gauss, G). Unless $B \gg 30$ G (which is strongly disfavored by multiple observational constraints; Sharma, Quataert, & Stone 2007), the observed NIR spectrum *requires* non-thermal particles having energies well above that associated with the quiescent 100 GHz brightness temperature of $\simeq 3 \times 10^{10}$ K (Bower et al. 2006) (i.e., $\gamma \sim 10$). This is why we focus on modeling the non-thermal distribution function in this paper. Moreover, we also focus largely on synchrotron radiation as the source of the flaring in both the NIR and X-rays; the alternative possibility, that the X-rays are produced by inverse-Compton upscattering of lower energy photons, is disfavored, at least for luminous flares (Dodds-Eden et al. 2009; see also §3).

Because there is no first-principles understanding of what generates the flares from Sgr A* (i.e., the source of particle acceleration), our models are necessarily somewhat phenomenological. In an attempt to go beyond phenomenological modeling, we also present results from time-dependent magnetohydrodynamic (MHD) simulations of accretion disks in which magnetic reconnection close to the last stable orbit dissipates magnetic energy in a manner similar to that required to explain the observed flaring from Sgr A* (see Yuan et al. 2009 and Ding et al. 2010 for related ideas).

The remainder of this paper is organized as follows. Section 2 provides a concrete physical model for the flaring from Sgr A*, motivated by "flares" in MHD simulations of accretion disks; some of the input parameters in our lightcurve models are motivated by these numerical results, but the lightcurve models are more general and are independent of the simulation results. Section 3 presents our calculations of the time-dependent evolution of the non-thermal electrons and the resulting lightcurves in different wavebands. We summarize and discuss the implications of our work in Section 4.

## 2. FLARING IN ACCRETION DISK SIMULATIONS

Global MHD simulations of RIAFs have been carried out extensively in the last decade, both with non-relativistic (e.g., Armitage 1998; Kudoh, Matsumoto, & Shibata 1998; Stone & Pringle 2001; Hawley & Balbus 2002; Igumenshchev, Narayan, & Abramowicz 2003) and relativistic codes (e.g., De Villiers, Hawley, & Krolik 2003; McKinney 2006; Mościbrodzka et al. 2009; Fragile & Meier 2009). The basic structure of such flows consists of a thick dense disk (moderately magnetized with ratio of gas pressure to magnetic pressure $\beta \equiv 8\pi p/B^2 \sim 1-100$) surrounded by a hot magnetically dominated corona, with jets launched near the last stable orbit (the efficiency of jet production depends on the imposed magnetic geometry and the spin of the black hole; Beckwith, Hawley, & Krolik 2008). In addition, the plasma within the last stable orbit near the equator (i.e., "disk" material) is in radial free-fall and becomes magnetically dominated ($\beta \ll 1$; e.g., Stone & Pringle 2001). Here we suggest that the observed flaring from Sgr A* may be due to magnetic reconnection close to the plunging region near the last stable orbit.

Magnetic reconnection has been invoked in past work (e.g., Markoff et al. 2001; Baganoff et al. 2001; Liu et al. 2004), but here we show explicitly that reconnection events can occur in numerical simulations of hot, magnetized accretion flows. The simulations that we present are essentially identical to Stone & Pringle (2001); Sharma, Quataert, & Stone (2007). The only difference is that we look for and find flaring activity, driven by reconnection, at short timescales comparable to the orbital period at the last stable orbit. Previous simulations were largely focused on the time averaged structure of the accretion flow, while here we study short timescale reconnection events.

### 2.1. *Numerical Setup and Initial Conditions*

The numerical methods and initial conditions used here are described in detail in Sharma, Quataert, & Stone (2007) (and references therein) so we only briefly review the key points. As in our previous work, we have carried out two-dimensional non-radiative accretion flow simulations in spherical $(r, \theta)$ geometry using the widely used ZEUS-MHD code (Stone & Norman 1992a,b). We solve the standard equations of MHD in the pseudo-Newtonian potential of Paczynski & Wiita (1980): $\Phi = -GM/(r - r_g)$, where $r_g = 2GM/c^2$. Although we are not using a conservative code, we can capture a reasonable fraction of the dissipated magnetic energy using an explicit resistivity of the form (Stone & Pringle 2001)

$$\eta = \eta_0 dr^2 \frac{|\nabla \times B|}{\sqrt{4\pi\rho}}, \qquad (1)$$

with $\eta_0 = 0.15$. The resistive terms in the induction and internal energy equations are included using the method of Fleming, Stone, & Hawley (2000). The plasma in the simulations with resistivity is hotter (especially in regions with high current density) than in the simulations without resistivity because most of the dissipated magnetic energy is captured as heat; the dynamics and flaring are, however, essentially identical.

There is no physics in our simulations that picks out an absolute density scale or spatial scale. To express the numerical results in units relevant for observations of Sgr A*, we present all of our results for an $M_{\rm BH} = 4 \times 10^6 M_\odot$ black hole with a time averaged accretion rate of $\dot M_{\rm in} = 10^{-8} M_\odot {\rm yr}^{-1}$; the latter is chosen for consistency with the measured Faraday Rotation (Bower et al. 2003; Marrone et al. 2007; Sharma, Quataert, & Stone 2007).

We use a $120 \times 88$ logarithmic grid (in both $r$ and $\theta$) extending from $2r_g$ to $800r_g$. The resolution is $\Delta\theta \sim \Delta r/r \approx 0.05$. The boundary conditions are the same as in Sharma, Quataert, & Stone (2007). Strict outflow boundary conditions are applied at both the inner and outer radial boundaries (plasma is not permitted to enter the computational domain); scalar quantities and $\theta$ & $\phi$ components of vectors are copied from the closest active zones. The magnetic stress is required to be positive ($B_r B_\phi \leq 0$) at the inner radial boundary so that matter is not pulled into the computational domain from the inner boundary. Reflective boundary conditions are applied at $\theta = 0, \pi$ with $B_r$ copied, and $B_\theta$ & $B_\phi$ reflected.

The simulations initialize a dense, constant specific angular momentum torus as in Stone & Pringle (2001); the initial density maximum of the torus is at $200r_g$. The calculations that we focus on initialize a single poloidal magnetic loop threading the initial torus, with field lines aligned with the surfaces of constant density; we will also briefly mention results for simulations with two initial magnetic loops in the dense torus (see Fig. 8 of Sharma, Quataert, & Stone 2007). The results of disk simulations remain sensitive to the initial magnetic field geometry, even at late times (Sharma, Quataert, & Stone 2007; Beckwith, Hawley, & Krolik 2008).

### 2.2. Simulation Results

The initial dense torus becomes unstable to the magnetorotational instability (MRI; Balbus & Hawley 1991), causing mass to flow in toward the central black hole in form of a thick (since the disk is non-radiative) accretion disk that remains threaded by the initial magnetic field. Because we initialize a coherent poloidal magnetic field loop threading the torus, a current sheet is formed in the equatorial region. To search for flaring, we first wait for sufficient time that a quasi-steady accretion flow is established at small radii ($\approx 2$ orbits at the initial density maximum). We then search for sudden temporal and spatial changes in physical quantities like the magnetic field strength and temperature, varying the spatial and temporal scales over which we analyze the results. For example, we analyzed the volume averaged magnetic field strength and temperature in the inner $6r_g$ of the simulation domain, sampled every $\approx 8$ minutes (the orbital timescale at the last stable orbit is $\approx 17$ minutes for Sgr A*). We find three magnetic reconnection events in this region in an interval of 4000 minutes. These are identified by a sudden decrease in the volume averaged magnetic field strength (from $\approx 60$G to $\approx 20$G) and an associated increase in temperature (by 10-100 times the quiescent value of $\approx 5 \times 10^{11}$ K). The temperature in these MHD simulations should be thought of as the ion temperature; the electron temperature is likely somewhat smaller. Thermal heating and nonthermal particle acceleration of both electrons and ions occurs during magnetic reconnection in solar flares (Lin et al. 2003). However, the quantitative details of these processes are not well understood, and as a result we will treat electron acceleration phenomenologically.

The reconnection event – identified with a sudden drop in magnetic field strength in the inner regions of the accretion flow and a simultaneous rise in the temperature – happens over $\approx 20$ minutes, the dynamical timescale in this region; this is also comparable to the typical duration of the X-ray and NIR flares. After the plasma is heated, mass, energy, and magnetic flux are expelled outwards as the over-pressured plasma expands and is pulled out by magnetic tension; once the hot, expanding plasma escapes, the accretion flow starts to build up again. This whole process takes $\approx 150$ minutes after which accretion is again in the quasi-steady state. The three "flares" we have identified are all qualitatively similar to each other. We describe one in more detail below. These flares are different from the variability expected from turbulent fluctuations in density and magnetic field strength in approximately the same region, which can produce fluctuations in the mm emission (Chan et al. 2009; Dexter, Agol, & Fragile 2009).

Figure 1 shows two-dimensional snapshots of the magnetic field strength and the magnetic field unit vectors during one of the flares, from the beginning at $t = 0$ to when quasi-steady accretion is re-established at $t = 152.9$ minutes. Figure 2 shows the plasma temperature and velocity unit vectors at the same times. The quiescent magnetic field is split-monopolar with the field reversing in the equator. This is a consequence of the field we initialize at larger radii; it is not guaranteed that the disk will in fact have such a magnetic field.

The first panel in Figure 1 shows the beginning of the flare when the magnetic energy has started to decrease; this magnetic dissipation is accompanied by a significant increase in the temperature in the central few $r_g$ (first panel of Fig. 2). By the second panel in Figure 1 at $\simeq 16$ min, the magnetic energy in the central region has decreased significantly and the equatorial current sheet in the disk (of low magnetic energy) is pushed out due to magnetic reconnection. This is because the hot supervirial plasma created near the last stable orbit is expanding outwards, as indicated by the velocity vectors in the top two panels of Figure 2. Most of the expanding material takes the path of least resistance and flows out along the "jet" at intermediate latitudes; note that this outflow is present at all times, but is significantly stronger during and just after the reconnection event. Once the hot, super-virial plasma leaves the inner region (bottom two panels in Figs. 1 & 2), the split-monopolar magnetic field geometry is re-established by accretion of plasma from larger radii. This last snapshot represents the 'quiescent' quasi-steady disk structure.

To test the sensitivity of these magnetic reconnection events to the initial magnetic geometry, we also studied an initial field geometry consisting of two poloidal field loops in the initial torus, with a net radial field in the equatorial plane (Fig. 8 of Sharma, Quataert, & Stone 2007). The structure of the resulting turbulent disk is somewhat different. The disk is thicker and less dense because of the magnetic pressure provided by the strong toroidal magnetic fields in the disk mid-plane (see Figures 9 & 10 of Sharma, Quataert, & Stone 2007). Magnetic reconnection events also occur near the last stable orbit for this initial magnetic field geometry, but they are not as dramatic and well-defined as in the case of a single initial loop. The reconnection events occur in the current sheets sandwiching the equatorial accretion disk; they are again accompanied by sudden heating of the inner regions due to magnetic dissipation. However, in this case the energy release is not sufficient to completely disrupt the quiescent accretion flow. An analogy to solar activity may be useful: in the simulations we focus on here (Figs 1 & 2), reconnection in the central region leads to a strong outflow qualitatively similar to a coronal mass ejection (CME) from the sun. In the case of two initial poloidal field loops, the reconnection events we find are more akin to true solar flares; i.e., magnetic energy is dissipated locally, but there is no large-scale outflow.

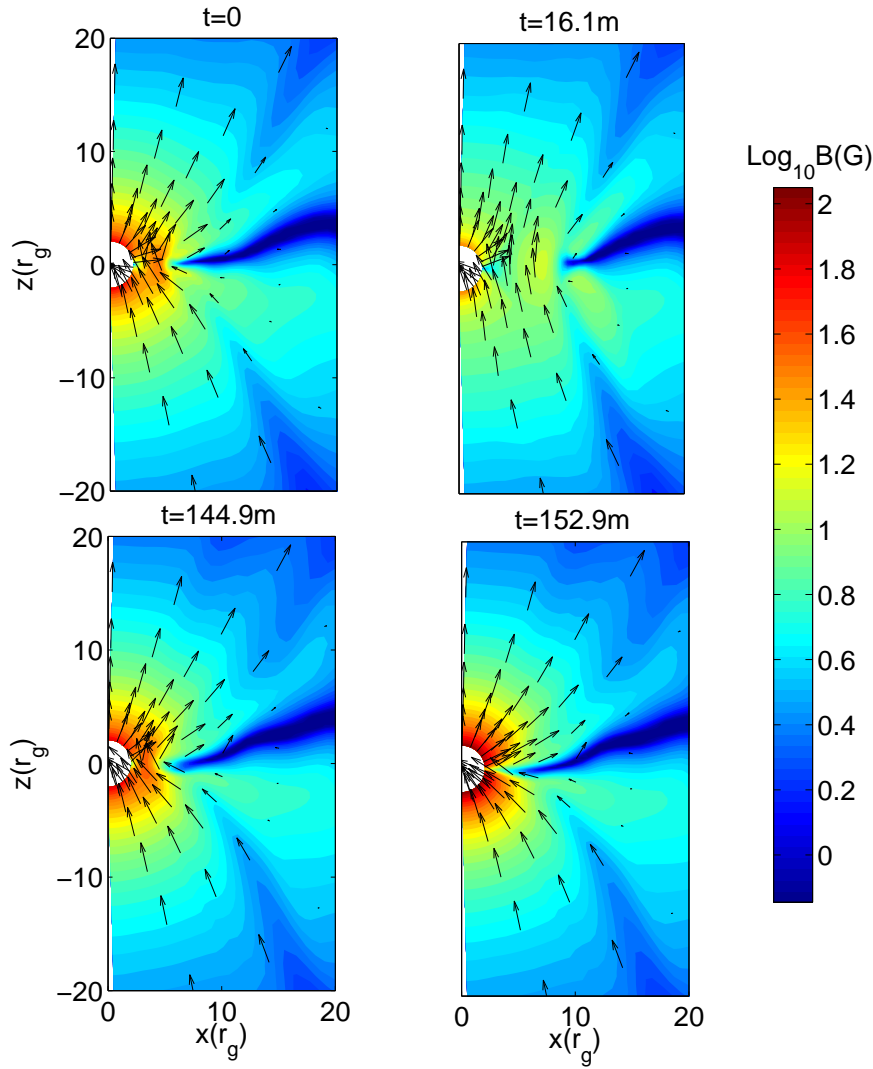

FIG. 1.— Two-dimensional contour plots of magnetic field strength (in G) at different times during one of the magnetic reconnection events identified in our simulations. Arrows denote the projection of the magnetic field unit vectors. Time $t = 0$ corresponds to the beginning of the magnetic reconnection event, and at $t = 152.9$ minutes the quasi-steady magnetic field structure and the equatorial current sheet are re-established. The snapshot just before t=0 looks similar to the quiescent state at $t = 152.9$. The magnetic field strength and times are plotted assuming $M_{\rm BH} = 4 \times 10^6 M_\odot$ and $\dot{M}_{\rm in} = 10^{-8}\ M_\odot\ {\rm yr}^{-1}$, as is reasonable for Sgr A*.

It is important to stress that the simulations presented here (Figs. 1 & 2) do *not* demonstrate that magnetic reconnection and flaring near the last stable orbit necessarily occur in MHD disk simulations (let alone in Sgr A*!). Rather, we demonstrate a weaker point: for disks with relatively coherent poloidal magnetic fields, simulations show magnetic reconnection events in which magnetic dissipation leads to localized heating and expansion of plasma. The oppositely directed magnetic fields required for these flares are generated by dragging in the coherent field lines from larger radii. Similar reconnection events are also present in axisymmetric MHD simulations of spherical accretion (Figs. 15 & 16 of Sharma, Quataert, & Stone 2008) and appears to be a generic feature of MHD accretion when inflow brings together oppositely directed field lines.

Future three-dimensional MHD simulations are required to explore in more detail the conditions under which this kind of reconnection can occur. Three-dimensional simulations, (e.g., Hawley & Balbus 2002) do show an overall flow structure that is quite similar to two-dimensional simulations (e.g., Stone & Pringle 2001). However, the critical issue here is related to the dynamo generation of large-scale magnetic fields in disks, which is still poorly understood. Because of these uncertainties, it is also not possible for us to reliably determine the statistics of reconnection events in the simulations for comparison to observations. However, the consistency with the energetics and timescales of the observed flares is encouraging and it seems plausible that the X-ray/IR flares observed in Sgr A* are due by reconnection, qualitatively similar to what is seen in our MHD simulations.

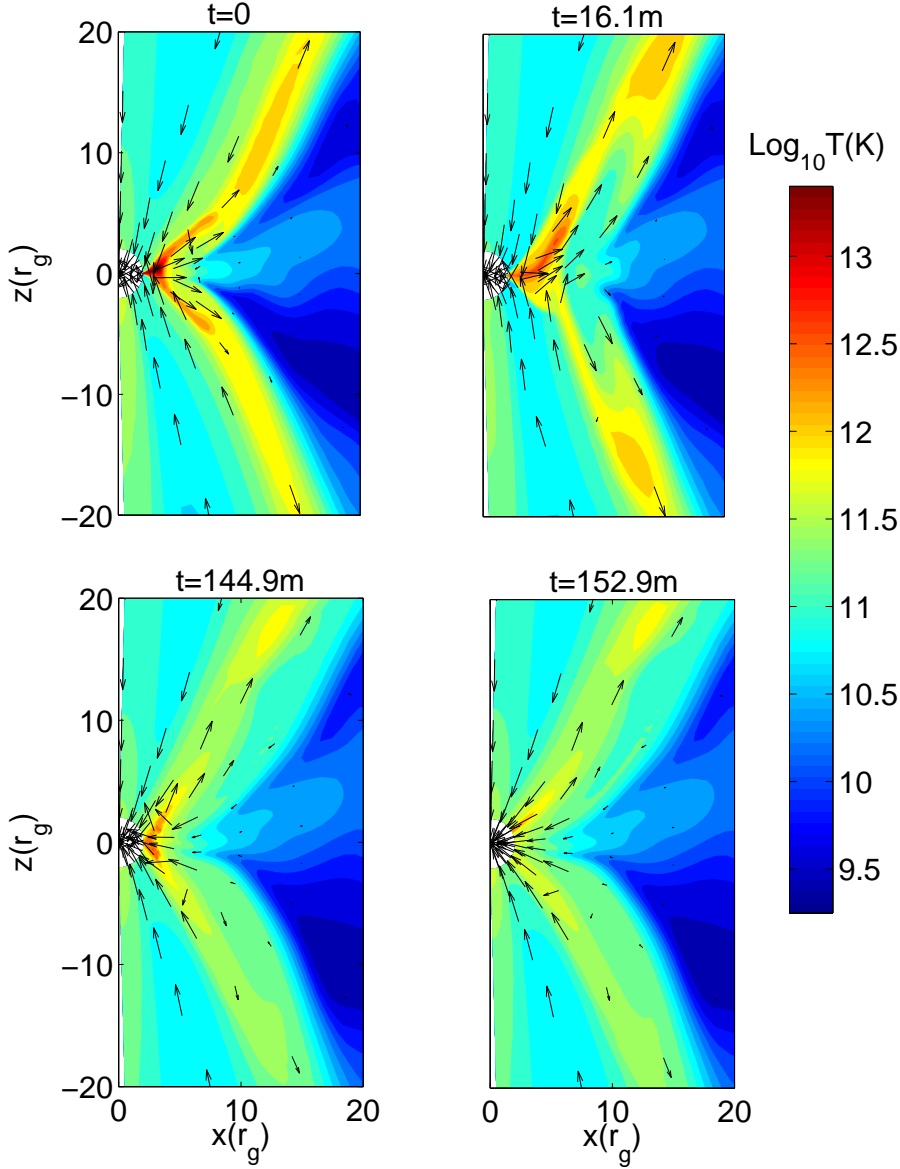

FIG. 2.— Two-dimensional contour plots of temperature (in K) at different times during the magnetic reconnection event shown in Figure 1. Arrows show the projection of velocity unit vectors. Starting at the beginning of the flare at $t=0$, hot, over-pressured plasma expands outwards at mid-lattitudes at close to the speed of light. A quasi-steady accretion flow reforms at $t=152.9$ minutes. The snapshot just before t=0 looks similar to the quiescent state at $t=152.9$. The temperature is independent of the black hole mass and the accretion rate in these RIAF simulations.

In the next section, we present a simple non-thermal electron acceleration model to explain the observed flares from Sgr A* in the context of both quasi-stationary heated plasma (corresponding to a mild flare) and expanding plasma (a CME-like flare).

### 3. LIGHTCURVE MODELING

In order to understand the time-dependent emission from nonthermal particles in Sgr A*, we have developed a simple model to describe the evolution of a transiently heated electron population. Some of the parameters we consider in these models are motivated by the simulations described in the previous section, but our lightcurve modeling is more general and does not depend in detail on the source of the energetic particles.

We first investigate general properties of time-dependent models given different assumptions about the dominant energy loss mechanism (for example, synchrotron cooling or adiabatic expansion). We then apply these models to the observed properties of the very bright and very high quality simultaneous NIR/X-ray flare from Sgr A* that was observed on April 4, 2007 (Dodds-Eden et al. 2009; Porquet et al. 2008; Trap et al. 2009; Yusef-Zadeh et al. 2009). The April 4, 2007 event, observed with NACO (VLT) at $3.8\mu m$ and with XMM-Newton at 2-10 keV, was sufficiently bright that detailed lightcurves were obtained simultaneously at multiple wavelengths, giving us an unprecedented chance to use the observations to explore the acceleration and cooling physics in Sgr A*. Key properties of this luminous flare include (Dodds-Eden et al. 2009, see also Figures 5, 9 and 10 later in this paper):

- The NIR and X-ray lightcurves are both relatively symmetric and peak simultaneously (to within $\approx 5$ minutes), and yet
- the $3.8\mu m$ rise and decay times are slower than in the X-ray, and the FWHM of the $3.8\mu m$ lightcurve is about twice that of the X-ray
- the NIR lightcurve shows dramatic substructure that is not present at the same level in the X-ray lightcurve (see Fig. 9)
- the X-ray flare has an average spectral index ($\nu L_\nu \sim \nu^\alpha$) of $\alpha_X < 0$ with 90% confidence (Porquet et al. 2008) while the NIR flux together with an upper limit at $11.88\mu m$ favors a spectral index $\alpha_{L'} > 0$. Previous NIR observations find $\alpha_{L'} \sim 0.4$ for luminous flares (Hornstein et al. 2007).

In Dodds-Eden et al. (2009) we proposed a "cooling break" synchrotron model to explain the average spectral properties of this flare and, by extension, all luminous flares from Sgr A*, which have similar properties (see Kardashev et al. 1962 for the basic theory of cooling breaks). In this model, both the NIR and X-ray emission are synchrotron emission, but the effect of synchrotron cooling is to produce a spectral break between the NIR and X-ray with $\Delta\alpha = 0.5$. The observed NIR and X-ray spectra are consistent with such a spectral break. The cooling break synchrotron model forms the basis of the models we explore here. In particular, observational and theoretical considerations strongly disfavor Inverse Compton (or Synchrotron Self-Compton) emission as the source of the observed X-ray flare (Dodds-Eden et al. 2009). We also explore the implications of our models for longer wavelengths (submm-radio), for which there was also data taken on April 4, 2007; in particular, a $\sim 0.4$ Jy flare was seen at 43 GHz with a duration of 100 minutes and a delay of $\sim 6$ hours relative to the NIR/X-ray event (Yusef-Zadeh et al. 2009).

### 3.1. Numerical Model

We calculate the synchrotron emission from an evolving population of electrons. The electron distribution function $N_e(\gamma,t)$ (units: number of electrons per unit Lorentz factor) evolves according to the following continuity equation (Blumenthal & Gould 1970):

$$\frac{\partial N_e(\gamma,t)}{\partial t} = Q_{\rm inj}(\gamma,t) - \frac{\partial [\dot\gamma N_e(\gamma,t)]}{\partial \gamma} - \frac{N_e(\gamma,t)}{t_{\rm esc}} \quad (2)$$

where $Q_{\rm inj}(\gamma,t)$ is the rate at which electrons of Lorentz factor $\gamma$ are injected at time $t$, the second term takes into account the redistribution of electrons in energy (i.e., cooling processes), and the third term allows the possibility that electrons may escape the region, with $t_{\rm esc}$ the escape timescale. We take $Q_{\rm inj}(\gamma,t)$ to be a power-law in $\gamma$: $Q_{\rm inj}(\gamma,t) = c_{\rm inj}(t)\gamma^{-p}$ with the normalization of the power law a function of time, $c_{\rm inj}(t)$, which we call the injection profile. The injected energy distribution has an exponential cutoff at $\gamma_{\max}$, which we set to $10^6$ in our calculations (the exact value of $\gamma_{max}$ is not important), and a minimum energy of $\gamma_{\min} \sim 10$. This choice of minimum energy assumes that the electrons are accelerated out of the thermal population of electrons, which have $T_e \sim 3 \times 10^{10}$ K in many models (e.g., Sharma et al. 2007a), and which emit in the sub-mm (e.g., Yuan et al. 2004). Below $\gamma_{\min}$ the injected distribution breaks to a slope of $p = -2$, as would a thermal distribution.

The $\dot\gamma$ term in equation (2) accounts for energy loss. We consider both synchrotron cooling and adiabatic expansion:

$$\dot\gamma = \dot\gamma_{\rm synch} + \dot\gamma_{\rm ad}. \quad (3)$$

For synchrotron cooling, $\dot\gamma_{\rm synch} = -\gamma/t_{\rm cool}$ where $t_{\rm cool} = 7.7462 \times 10^8/(\gamma B^2)$ for an isotropic distribution of pitch angles (Rybicki & Lightman 1979). For adiabatic expansion, $\dot\gamma_{\rm ad} = -\gamma\, d\ln R/dt$ where $R(t)$ is the radius of the volume of interest (approximated as a sphere for simplicity). We express adiabatic losses in terms of $R$ rather than the electron number density $n_e$ because the latter can change by injection/escape which do not, however, modify the energy of the particles.

The electron distribution function and lightcurves depend on the value of $B$ and on how both $B$ and $R$ change with time. For an adiabatically expanding plasma, the magnetic field $B(t)$ will be affected by the expansion, in addition to $n_e$ and $\gamma$. For a fixed radial magnetic field, magnetic flux conservation implies $B \sim 1/R(t)^2$, but the dependence can be more general depending on the magnetic field geometry.

The synchrotron emission is calculated at each time given the instantaneous electron energy distribution using formulae from Rybicki & Lightman (1979). We calculate the emission coefficient from

$$j_\nu = \frac{1}{4\pi}\int_1^\infty n_e(\gamma)\langle P_e(\gamma,\nu,\phi)\rangle d\gamma \quad (4)$$

where we approximate the pitch-angle ($\phi$) averaged spectral power emitted by a single electron $\langle P_e(\gamma,\nu,\phi)\rangle$ using

$$P_e(\gamma,\nu,\phi) = \frac{\sqrt{3}q^3 B}{mc^2} F\left(\frac{\nu}{\nu_{syn}(\gamma,\phi)}\right)\sin\phi \quad (5)$$

evaluated at $\phi = \arcsin(\pi/4)$ (which is close to the true pitch angle averaged spectrum and much faster to evaluate). The angle-averaged value of the critical synchrotron frequency is $\langle \nu_{\rm syn}(\gamma,\phi)\rangle = 3qB\gamma^2/(16m_e c)$, where $q$, $m_e$ and $c$ are the electron charge, mass, and the speed of light. The function $F(x) = x\int_x^\infty K_{5/3}(\xi)d\xi$ in equation 5 describes the shape of the spectrum.

We calculate the absorption coefficient from

$$\alpha_\nu = \frac{c^2}{8\pi\nu^2 mc^2}\int_1^\infty n_e(\gamma)\left(\frac{2P_e(\gamma)}{\gamma} + \frac{dP_e(\gamma)}{d\gamma}\right)d\gamma \quad (6)$$

and then the resultant emission, assuming a homogeneous sphere of radius $R$ (Gould 1979), using

$$\nu L_\nu = 4\pi^2 R^2 \frac{\nu j_\nu}{\alpha_\nu}\left(1 + \frac{\exp(-2\alpha_\nu R)}{\alpha_\nu R} - \frac{1-\exp(-2\alpha_\nu R)}{2\alpha_\nu^2 R^2}\right). \quad (7)$$

We checked our numerically computed spectra against the analytical equations of Gould (1979) and Marscher

(1983). Note that if the emission is optically thin, the luminosity only depends on the total number of accelerated electrons $N_e \sim n_e R^3$. There is thus a degeneracy between the number density of accelerated particles $n_e$ and the size of the flaring region $R$. By contrast, if optical depth effects are important, which is the case for the radio emission, this degeneracy is broken.

The end result of our calculation is the self-consistently determined time-dependent synchrotron SED, given the following possibly time-dependent input parameters: magnetic field $B(t)$, particle injection rate $c_{\rm inj}(t)$, particle index of the injected electron distribution $p$ (which we take to be constant in time), and radius of the emission region $R(t)$. In Section 3.5 we also calculate the inverse Compton scattered spectrum, which provides an independent constraint on the size of the emission region; we use the prescriptions of Blumenthal & Gould (1970).

### 3.2. Lightcurves

Figure 3 shows model NIR (L'-band; 3.8 $\mu$m) and X-ray (4.1 keV; $10^{18}$Hz) lightcurves and spectral indices, for three different assumptions about the dominant electron energy loss mechanism: synchrotron cooling, escape, and adiabatic expansion (corresponding time-dependent SEDs for the different cooling mechanisms are shown in Figure 4). The parameters of these models are given in Table 1, where the models are parameterized by the total number of electrons at the peak of the X-ray flare; the particle index, $p$, the value of the magnetic field at the peak $B_{\rm peak}$ (i.e., $B$ at 56 minutes); the escape time $t_{\rm esc}$; and the (constant) expansion velocity, $v_{\rm exp}$, in units of $R_i$/hr where $R_i$ is the initial radius (expansion begins at $t = 0$). The actual value of $R_i$ does not matter here because both the 3.8$\mu$m and X-ray emission are optically thin (hence we did not apply optical-depth effects in these calculations). The injection profile of the accelerated particles is the same in all of the models: a Gaussian with a FWHM of $\simeq 27.5$ min. In each calculation, the particle index $p$ was adjusted (see Table 1) so that the peak IR luminosity was comparable to the peak X-ray luminosity; larger $p$ would lead to a lower X-ray luminosity relative to the IR luminosity, and vice-versa, but otherwise the value of $p$ does not change any of our conclusions about the NIR and X-ray flares.

Figure 3 shows that the X-ray and NIR lightcurves behave very differently for different model parameters. The X-ray lightcurve is almost independent of the model details, while the NIR lightcurve is much more sensitive. This difference is due to the very different synchrotron cooling timescales for electrons emitting in the IR and X-ray. The synchrotron cooling timescale for an electron emitting at a given frequency $\nu$ is given by

$$t_{\rm cool} \simeq 8 \left(\frac{B}{30\,\rm G}\right)^{-3/2} \left(\frac{\nu}{10^{14}\,\rm Hz}\right)^{-1/2} \text{ min.} \quad (8)$$

In the X-ray, the synchrotron cooling timescale is almost always much shorter than the injection timescale. In that limit, the electrons radiate all of the energy they are supplied (via injection) and the emission is independent of the precise values of $\gamma$, $B$, $R$, etc. This is why the X-ray lightcurve closely follows the rate of electron injection in Figure 3, and is independent of the model details. By contrast, the L' emission, for most of the flare duration, occurs near or below the cooling break – the frequency at which the synchrotron cooling timescale is comparable to the injection time (which is expected to be of order the dynamical time). As a result, the NIR emission is sensitive to the details of the model. The general trend is the same regardless of the precise energy loss mechanism. Models with longer "cooling" times (be it via synchrotron, escape, or adiabatic losses) produce longer duration flares in the NIR, particularly when the cooling time is longer than the injection time. This is because the particle energy builds up initially and is then released over a longer period of time. These longer duration NIR lightcurves are also always delayed with respect to the peak of the injection and the peak of the X-ray lightcurve.

The results in Figure 3 are not consistent with some of the observed properties of the April 4, 2007 flare (see Figures 5, 9 and 10 for the observed data). For example, model L'-band lightcurves with the same duration as the April 4, 2007 flare ($\simeq 54$ min) are typically delayed from the X-ray lightcurve by 10-15 minutes, longer than that observed. Of the three different energy loss mechanisms, the adiabatic model does better than the other two in matching the longer duration and short delay (compare, e.g., the $B = 10$ G synchrotron cooling model, with a 19 min delay and 68 min duration, with the $v_{\rm exp} = 0.05 R_i/$hr adiabatic cooling model, which has a 9 minute delay and 55 min duration). The adiabatic expansion models have a shorter delay because the magnetic field decreases with the expansion ($B \propto 1/R^2$): this decreases the NIR emissivity of the electrons with time, and allows the lightcurve to peak earlier than it would due to the other cooling/loss processes.

The asymmetric NIR lightcurves in Figure 3 are also inconsistent with the April 4, 2007 flare, for which the lightcurves are reasonably symmetric and simultaneous and yet have different durations. A related problem is to understand the *rising phase* of the lightcurve. The slow rise of the NIR lightcurve compared to the X-ray lightcurve in the flares from Sgr A* cannot be explained by solely invoking a cooling mechanism, as is demonstrated in Figure 3, but requires a reduction in emissivity. If is, of course, possible that the discrepancies between the simple models in Figure 3 and the observed flares from Sgr A* are related to our overly simplistic treatment of particle acceleration. For example, lower energy (IR emitting) electrons need not have the same injection profile as higher energy (X-ray emitting) electrons. We do not explore this in detail, but instead focus on the possibility that changes in the magnetic field during flares strongly influence their observed properties.

If we define an "injection timescale" as $t_{\rm inj} = N_e(\gamma, t)/Q_{\rm inj}(\gamma, t)$ then the time-dependent position of the cooling break is given by

$$\nu_{\rm cool,syn} \approx \frac{10^{14}}{(p-1)^2} \left(\frac{B}{30\,\rm G}\right)^{-3} \left(\frac{t_{\rm inj}}{10\,\rm min}\right)^{-2} \text{ Hz.} \quad (9)$$

The relative luminosities in the X-ray and NIR depend on the position of the cooling break, which in turn depends on the injection timescale and the magnetic field strength. For example, for a fixed X-ray luminosity, a higher frequency cooling break corresponds to a lower NIR synchrotron luminosity (and vice-versa).

If the magnetic field were to *decrease* during the flare,

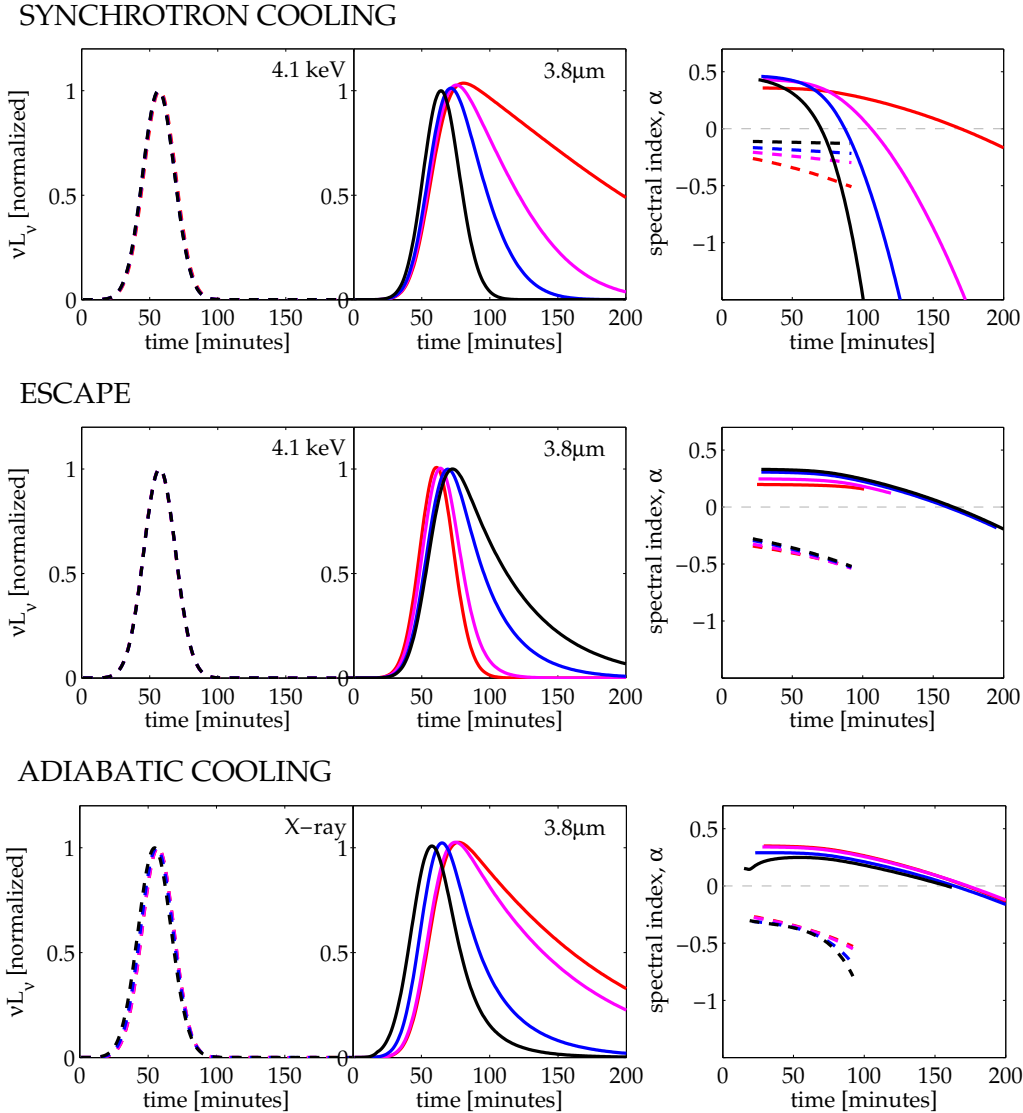

Fig. 3.— X-ray and NIR (L'-band) lightcurves and spectral indices for the flare evolution models given in Table 1. Both X-ray and L'-band lightcurves are normalized to the peak X-ray luminosity (the absolute luminosities are proportional to $N_e$ because the emission is optically thin). The spectral indices are only shown where the the corresponding lightcurve exceeds 1% of the peak flux. *Synchrotron cooling:* Black to red corresponds to $B = 30, 15, 10,$ and 5 G, respectively. *Escape:* Black to red corresponds to escape timescales of 60, 30, 10 and 5 min. *Adiabatic cooling:* Black to red corresponds to expansion velocities of 0.1, 0.05, 0.01 and 0.005 $R_i$/hr, where $R_i$ is the initial radius of the expanding plasma.

that would increase both the synchrotron cooling time and the cooling break frequency. An increasing cooling break frequency during the rising phase of the flare would in turn cause the NIR lightcurve to rise more slowly than the X-ray lightcurve, qualitatively consistent with observations. Another way to understand this is to note that the X-ray emissivity does not decrease if the magnetic field decreases because the synchrotron cooling time in the X-rays is shorter than the injection time. However, a decreasing magnetic field strength would decrease the emissivity in the NIR where the cooling time can be longer than the injection time. The decreasing $B$ required in this scenario is also consistent with the premise that magnetic energy dissipation generates the particle acceleration that produces the flare in the first place.

We now explore two ways in which the magnetic field might change during the flare: (i) a stationary solar-flare-like model and (ii) an expanding plasma model (analogous to a CME; see §2). In addition to the motivation for considering a varying $B$ given here, these two scenarios are also motivated by the numerical results discussed in §2.

### 3.3. *Quasi-stationary Flare Model*

Suppose that magnetic reconnection occurs somewhere in the inner regions of the accretion flow, as in Figure 1. In the region where the magnetic reconnection occurs the magnetic field decreases as magnetic energy is converted into the energy of accelerated particles. Here we consider a quasi-stationary flaring region like this that does not expand (so there are no adiabatic losses); this is applicable when the dissipated magnetic energy is smaller than the internal energy of the ambient accretion flow.

In Figure 5 we show several models in which electrons are injected into the emission region over a $\sim 30$ minute timescale, set by the duration of the X-ray flare. The models are compared to the NIR and X-ray lightcurves of the April 4, 2007 flare. Figure 5 shows three different models that all reproduce the NIR lightcurve reasonably well; the three models correspond to different values of $t_{\rm esc}$ (10, 30, & 200 min) and different $B(t)$. In this context, the escape of particles corresponds to accretion onto the black hole or escape in high speed outflows that do not emit significantly.

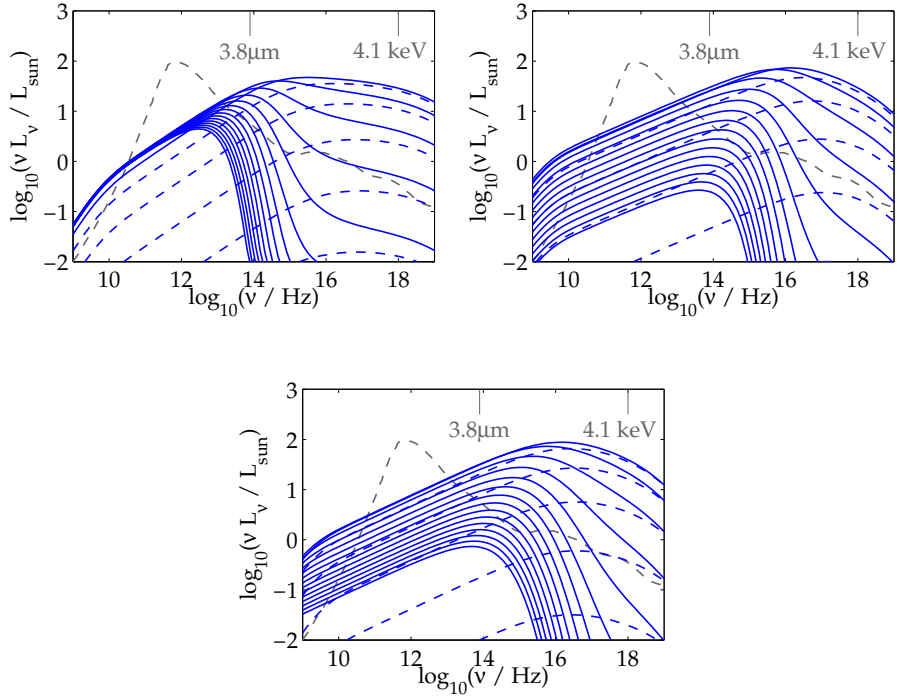

FIG. 4.— SED plots showing the evolution of the SED with time for models where the decline phase of the NIR lightcurve is dominated by synchrotron cooling, escape, or adiabatic cooling (corresponding to the lightcurves shown in blue in Figure 4). The dashed lines show the rising phase, while the solid blue lines show the evolution after the peak X-ray emission. The SED is plotted at 10 minute intervals. The quiescent model of Yuan et al. (2003) is plotted (dashed gray line) for reference. We did not apply optical depth effects in these models since we are focusing on the optically thin X-ray and NIR emission.

Figure 5 shows that if the magnetic field decreases from $\sim 40$ G at the beginning of the flare to $\sim$5-10 G at the peak of the flare, the simultaneity and symmetry of the NIR lightcurve, and the differences in duration of the NIR and X-ray lightcurves, are all reproduced reasonably well. For $t_{\rm esc} \gtrsim 30$ min, we find that $B$ must continue to decrease after the peak of the flare (or at least level off) in order to not overproduce the NIR flux at late times. By contrast, for short escape times $\sim 10$ min (blue lines in Figure 5), there are so few electrons around at late times that the magnetic field must increase again in order to produce the observed emission. This increase in $B$ at late times is reminiscent of the numerical simulations in Figures 1 and 2, in which the accretion flow returns to its quasi-steady state after the flare comes to an end. More generally, if the emission is dominated by particles that remain in the flare region, rather than expanding outwards in an outflow, then a decrease in $B$ followed by an increase is natural if magnetic energy is what generates the accelerated particles in the first place.

### 3.4. Expanding Plasma: a "Coronal Mass Ejection"

The change in magnetic field required to explain the NIR lightcurve in Figure 5 could be a consequence of outward expansion, rather than a local change in the magnetic field at a given position. Figure 6 shows an expansion profile $R(t)$ and $v(t)$ that can reproduce the properties of the NIR and X-ray flares, i.e., lightcurves like those in Figure 5 (we do not explicitly show the lightcurves and spectra for this model because they are similar to Fig. 5). The escape time is large for these models, $t_{\rm esc} = 1000$ min. Here we assume that the only change in the magnetic field is that produced by expansion, with $B \propto R^{-2}$. The expansion profile required to explain the NIR lightcurve is somewhat complicated, incorporating an acceleration, followed by deacceleration, followed again by an acceleration. This corresponds directly with the non-uniform variations in $B(t)$ required in Figure 5 (i.e. a sharp decrease in B during the rising phase of the flare, leveling off near the peak of the flare, thereafter decreasing once more). We suspect that in reality, the initial decrease in $B$ required to account for the NIR lightcurve (at $t \lesssim 50$ min) is due to a sudden loss of magnetic energy (rather than expansion) – the later ($t \gtrsim 50$ min) expansion in Figure 6 may be due to slow expansion of the blob with the jet, as seen in the numerical simulations (see Fig. 2). The multi-dimensional expansion of the heated plasma in the simulations cannot be captured by the simple one-zone model considered here; the complexity in the expansion profile $v(t)$ required in Figure 6 may be a consequence of the limitations of this simple model.

TABLE 1
Properties of L'-band and X-ray flares in the time-dependent synchrotron model

| Model | Remarks | $N_e(t_{\rm peak})$ [$\times 10^{45}$] | p | $B(t_{\rm peak})$ [G] | $t_{\rm esc}$ [min] | $v_{\rm exp}$ [$R_i$/hr] | $\Delta t_{\rm peak}$ [min] | fwhm$_{\rm X}$ [min] | fwhm$_{\rm L}$ [min] |
|---|---|---|---|---|---|---|---|---|---|
| *Synch cooling* | $B$ = const | 0.026 | 1.95 | 30 | 1000 | 0 | 2 | 27.5 | 28 |
| | | 0.057 | 2.03 | 15 | 1000 | 0 | 7 | 27.5 | 31 |
| | | 0.13 | 2.11 | 10 | 1000 | 0 | 19 | 27.5 | 68 |
| | | 0.81 | 2.27 | 5 | 1000 | 0 | 26 | 27.5 | 152 |
| *Escape* | $B$ = const | 7.2 | 2.59 | 5 | 5 | 0 | 4 | 27.5 | 29 |
| | | 3.9 | 2.49 | 5 | 10 | 0 | 7 | 27.5 | 32 |
| | | 1.7 | 2.37 | 5 | 30 | 0 | 12 | 27.5 | 44 |
| | | 1.2 | 2.32 | 5 | 60 | 0 | 16 | 27.5 | 59 |
| *Adiabatic cooling* | $v_{\rm exp}$=const & $B \propto 1/R^2$ | 4.3 | 2.43 | 5 | 1000 | 0.1 | 4 | 36 | 46 |
| | | 2.5 | 2.38 | 5 | 1000 | 0.05 | 9 | 32 | 55 |
| | | 1.1 | 2.30 | 5 | 1000 | 0.01 | 18 | 28 | 97 |
| | | 0.95 | 2.29 | 5 | 1000 | 0.005 | 20 | 27.5 | 113 |
| *Decreasing B* | $B(t)$ | 3.1 | 2.4 | 5.1 | 200 | 0 | 3 | 31 | 59 |
| | | 2.5 | 2.4 | 5.6 | 30 | 0 | 3 | 31 | 60 |
| | | 1.6 | 2.4 | 7.1 | 10 | 0 | 3 | 31 | 62 |
| *April 4, 2007 L'/X-ray flare* | | | | | | | 3.4 ± 1.2 | 27.4 ± 1.4 | 54 ± 4 |

Note. — Parameters and results for the different flare models shown in Figures 3 and 4 (*Synch cooling, Escape, & Adiabatic cooling*) & 5 (*Decreasing B*). $N_e(t_{\rm peak})$ shows the total number of accelerated particles at the time when the X-ray flare peaks that would produce a peak X-ray luminosity of $30L_\odot$; however, the value of $N_e(t_{\rm peak})$ does not influence the lightcurve shape (i.e., duration, delays) because the NIR and X-ray emission are optically thin. The electron power-law index is $p$ ($n(\gamma) \propto \gamma^{-p}$); $B(t_{\rm peak})$ denotes the magnetic field strength at the time when the X-ray flare peaks; $t_{\rm esc}$ and $v_{\rm exp}$ denote the escape timescale and expansion velocity, with $R_i$ the initial radius of the expanding plasma. $\Delta t_{\rm peak}$ is the delay between the peaks of the model X-ray lightcurve and L'-band lightcurve (the X-ray lightcurve always peaks first). fwhm$_{\rm X}$ and fwhm$_{\rm L}$ are the full width half maximum widths of the X-ray and L'-band lightcurves, respectively. The injection profile in all models has the same FWHM of 27.5 minutes. In the adiabatic cooling models expansion begins at $t = 0$. In all cases, the electron power-law index $p$ is chosen so that peak X-ray and L'-band luminosities are comparable. The last row shows the observed delay and lightcurve widths for the April 4, 2007 flare (Dodds-Eden et al. 2009), calculated from the best fit Gaussians to the lightcurves (the values differ slightly from those given in Dodds-Eden et al. 2009, where the fwhm was measured directly from the lightcurves).

We now consider models in which we fix the properties of the flare for the first ∼ 100 min to be the same as in Figure 5, since this is what is required to explain the IR and X-ray emission. The NIR/X-ray flare is assumed to be static, but it sets the initial conditions for subsequent expansion (i.e., $B \simeq 5$ G and $N_e \simeq 10^{45}$). These calculations are motivated in part by observations that suggest a delay between longer wavelength (radio-mm) variations and the simultaneous NIR/X-ray events (Yusef-Zadeh et al. 2006; Hornstein et al. 2007; Marrone et al. 2008; Yusef-Zadeh et al. 2008; Meyer et al. 2008; Eckart et al. 2008), where the delayed long wavelength emission might result from outflowing plasma (van der Laan 1966).

The additional parameters determining the properties of later timescale emission are the initial size of the emitting region (prior to expansion), $R_i$, and the expansion speed $v_{\rm exp}$. The initial size $R_i$ does not change the results of the NIR/X-ray emission because those wavelengths are optically thin (the flare emission only depends on the total number of accelerated particles $N_e$, not on $R_i$ and the electron density $n_e$ separately); however, lower frequency emission can become self-absorbed, which introduces a dependence on $R_i$.

Figures 7 and 8 shows lightcurves and time-dependent SEDs for models which fit the NIR/X-ray lightcurves of the April 4, 2007 flare, but in which we also initiated expansion at $t \approx 100$ minutes. Unlike in §3.3, here we set the escape time to $t_{\rm esc} = 2000$ min, so that no electrons escape on any timescales of interest; this is motivated by the fact that we are now following the thermodynamics of the expanding electrons. In Figure 7 we show the effects of considering different values of $R_i$ and $v_{\rm exp}$, while Figure 8 shows the effects of changing the power-law index of the injected electrons $p$, as well as the effect of different functional forms for $B(R)$, relaxing the assumption we made previously of expansion in a purely radial field ($B \sim 1/R^2$).

One prominent feature of the model lightcurves in Figures 7 and 8 is that there is in general no delayed mm emission; this is because the flare is optically thin at mm wavelengths (240 GHz). The low optical depth is partially due to the low initial magnetic field of 5 G (vs., say, 30 G) but more importantly it is because of the low electron densities. The only one of our models to show a delayed flare at mm wavelengths has $R_i = 0.1 r_g$ and thus an electron density of $n_e \simeq 10^{11}$ cm$^{-3}$. There are, however, independent observational arguments in favor of ambient densities $n_e \sim 10^7 - 10^8$ cm$^{-3}$ near Sgr A* (see Bower et al. 2003; Marrone et al. 2007). For $N_e \sim 10^{45}$ accelerated electrons inferred from the X-ray and NIR flares, we require $R_i \sim 1$-$2r_g$ to have $n_e = 10^7 - 10^8$ cm$^{-3}$ (see §3.5). Figures 7 and 8 show that for this size, the accelerated particles produce a simultaneous, rather than delayed, mm flare – this is a robust conclusion that is true for all expansion speeds, magnetic field geometries, etc. Increasing the initial magnetic field strength to ∼ 30

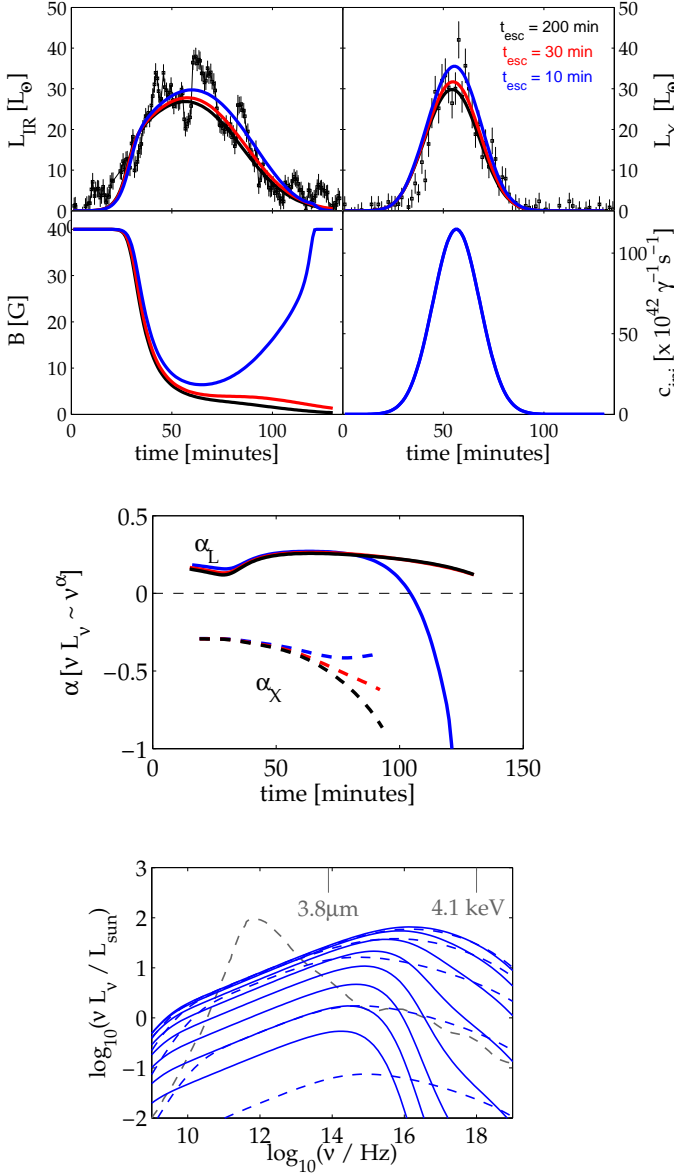

FIG. 5.— Stationary flare models with a decreasing magnetic field, envisioned to occur as a result of a magnetic reconnection in the inner regions of the accretion flow. We show results for three different escape timescales. The models are compared to the L'-band and X-ray data from the April 4, 2007 flare. *Top panel:* Model lightcurves, time dependent magnetic field and injection profiles. *Middle panel:* L'-band and X-ray spectral indices throughout the flare. *Lower panel:* The time-dependent SED for the $t_{\rm esc} = 30$ min model. The dashed lines show the rising phase, while the solid blue lines show the subsequent evolution. The SED is plotted at 10 minute intervals. The quiescent model of Yuan et al. (2003) is plotted (dashed gray line) for reference. No optical depth effects are included in these models since we focus on the optically thin NIR to X-ray emission.

G also does not qualitatively change this conclusion.[6]

In contrast to the optically thin mm emission, Figures

[6] For example, the following equations for the self-absorption frequency (for $\gamma_{min} = 10$), adapted from Gould (1979), show that upon substituting the values from Table 1 for both a $B \approx 5$ G model (the decreasing B model with $t_{esc} = 30$ min) and a $B = 30$ G model (a synchrotron cooling model) which produce the correct peak fluxes to match the NIR and the X-ray flare, the initial self-absorption frequency is only $\nu_{SSA} \approx 180 - 220$ GHz ($\lambda \approx 1.3$ mm) in both cases:

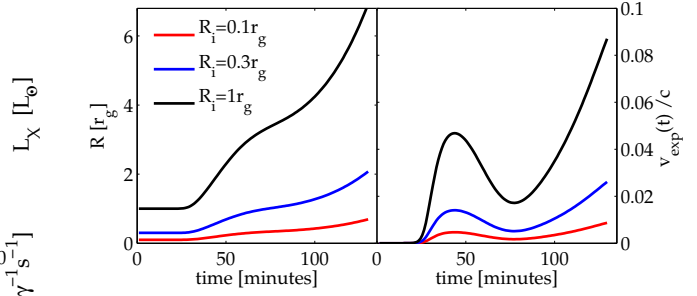

FIG. 6.— The expansion and velocity profiles $R(t)$ and $v_{\rm exp}(t)$ for a model which explains the NIR/X-ray lightcurves through adiabatic expansion; the magnetic field decrease is due solely to expansion with $B(R) \sim 1/R^2$. The lightcurves and magnetic field time dependence look very similar to the $t_{\rm esc} = 200$ min model of Fig. 5 so we have not shown them here. The expansion profile has an acceleration followed by de-acceleration, followed by another acceleration. The variation is such that $1/t_{\rm exp} = d \ln R/dt (= v/R)$ is the same for different $R_i$.

7 and 8 show that our models predict that there could be optically thick flares at 43 GHz delayed by $\sim 100 - 200$ min relative to the NIR and X-ray emission. Indeed, due to lack of coverage at 43 GHz until 250 min after the flare, we cannot rule out that such a delayed flare occurred for the April 4, 2007 event. However, the delayed flares in our models typically have peak fluxes of only $\sim 0.1$-0.25 Jy for initial sizes of 1-2 $r_g$, and would be barely noticeable.

The flux of delayed radio emission is largest under one of two conditions. First, it is larger if the magnetic field decreases more slowly as the plasma expands (e.g., if $B \propto R^{-1}$ instead of $B \propto R^{-2}$, or even $B \propto$ const, which corresponds to expansion in a purely vertical magnetic field geometry); see the left panel in Figure 8. Secondly, the radio flux is also larger if there is a significant population of lower energy electrons, which do not emit in the mm to X-ray, but can emit at lower frequencies. This is demonstrated explicitly in the right panel of Figure 8, which shows that the delayed radio flux is larger for larger values of the electron power-law index $p$. Large values of $p \gtrsim 2.6$, however, become inconsistent with the NIR and X-ray fluxes and spectral indices. In addition $p \gtrsim 2.6$ imply a simultaneous mm flare of $> 1$ Jy; such large simultaneous variations in the mm flux are not observed (see Fig 10 below).

The observed April 4, 2007 flare from Sgr A* was followed by an increase in the 43 GHz flux to $\sim 1.5$ Jy, $\sim 400$ min after the flare (Yusef-Zadeh et al. 2009). Our results in Figures 7 and 8 demonstrate that this increase in the radio flux cannot be due to immediate expansion of the particles that produced the NIR and X-ray emission. This does not, of course, rule out that the two different "flares" are causally connected. For example, the expanding plasma that produced the NIR to X-ray emis-for the $B \approx 5$G model (p=2.4),

$$\nu_{SSA} = 220 \text{GHz} \left(\frac{B}{5.6\text{G}}\right)^{0.677} \left(\frac{N_e}{2.5 \times 10^{45}}\right)^{0.323} \left(\frac{R}{1.5 r_g}\right)^{-0.645}$$

for the $B = 30$G model (p=1.95),

$$\nu_{SSA} = 180 \text{GHz} \left(\frac{B}{30\text{G}}\right)^{0.664} \left(\frac{N_e}{2.6 \times 10^{43}}\right)^{0.336} \left(\frac{R}{1.5 r_g}\right)^{-0.672}$$

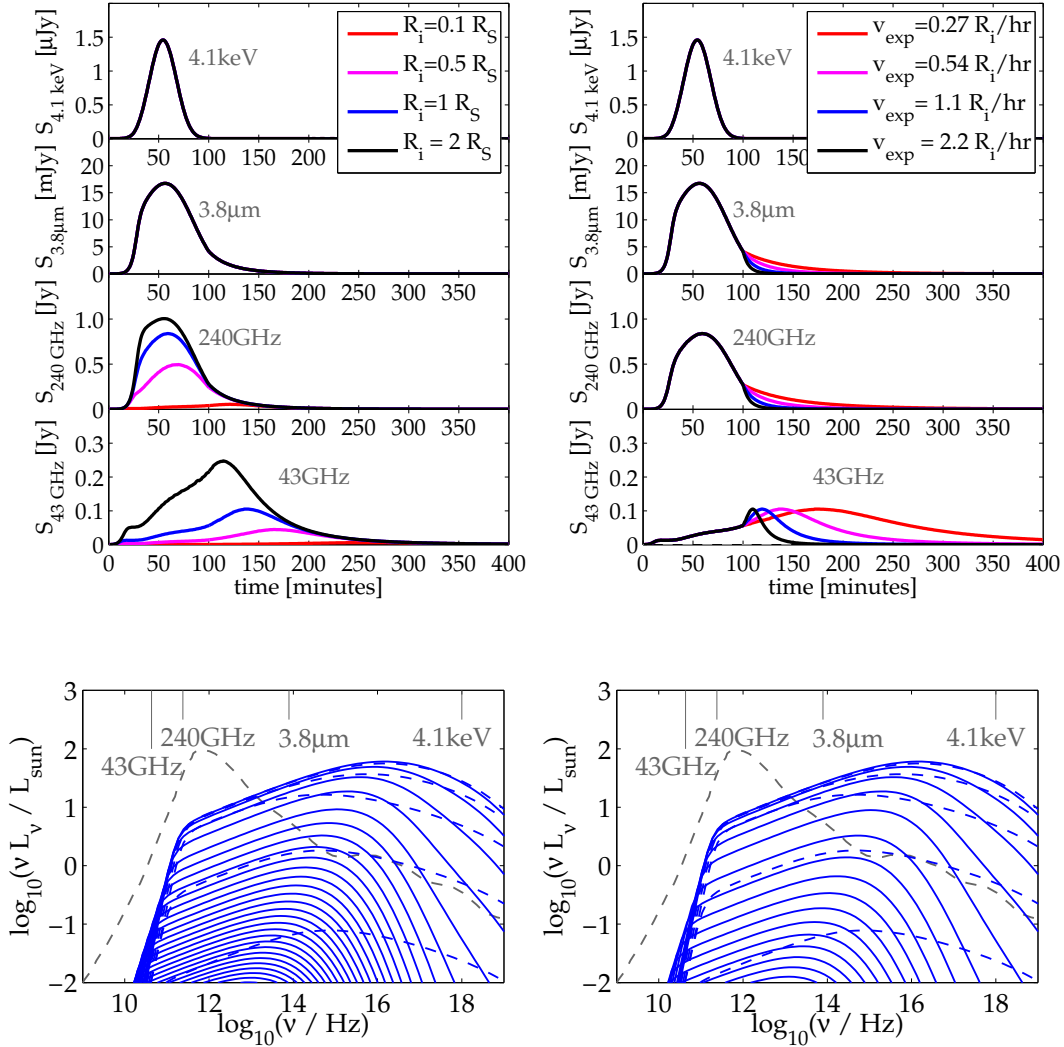

FIG. 7.— Adiabatic expansion models for radio-mm emission, including the effects of synchrotron self-absorption. The left panel shows the effect of varying $R_i$, the initial radius, the right panel the effect of different expansion speeds $v_{exp}$ (see Figure 1 for the effects of different $B(R)$ and $p$). In contrast to the previous figures we plot all lightcurves in $S_\nu$, for better comparison to the published mm/radio fluxes of the April 4, 2007 flare (the simultaneous 240 GHz flux was between ∼3-4 Jy; the 43 GHz radio flare, at a delay of 6 hours with a duration of 100 min, varied between ∼ 1.1 − 1.7 Jy – see Fig. 10). The model from Section 3.3 with $p = 2.4$, $B(t_{peak}) \simeq 5G$ and $t_{esc} = 10000$ min sets the initial conditions for the expansion, which begins at $t = 100$ min. When not otherwise listed in the legend the model has $v_{exp} = 0.54 R_i/$hr, $R_i = 1 r_g$, and $B \propto 1/R^2$. The SED plots show the evolution of the SED with time for the model shown in blue in the upper panel; the dashed lines show the injection phase, while the solid blue lines show the subsequent evolution. The quiescent model of Yuan et al. (2003) is plotted (dashed gray line) for reference. The time-dependent SED is plotted at 10 minute intervals.

sion could be reaccelerated as it moves outwards (e.g., via shocks), increasing the emission at later times above that predicted by our models.

The slow speeds required to produce delayed flares of ∼ 100 minutes are also much smaller than the escape speed close to the black hole, as has been noted in previous work (Yusef-Zadeh et al. 2006; Marrone et al. 2008). It is also apparent in comparing our model lightcurves with the observations (compare the right panel of Figure 7 with the 43 GHz observations in Figure 10) that these slow speeds are at odds with the relatively short duration of the observed 43 Ghz flare (compared to the length of the delay). It could be the case that the expansion of the plasma does not begin until well after the NIR and X-ray emission cease, such that a shorter duration lightcurve can be produced at longer delay. Indeed, the viscous time in the inner parts of the accretion disk is likely ∼ 10 dynamical times ∼ 200 min. Thus if the accelerated particles are not initially overpressurized (so that they do not expand on a dynamical time), they could be advected out in the 'quiescent' outflow after a few viscous times, producing – with some re-acceleration – delayed radio emission on approximately the correct timescale. However, it is clear that significant fine-tuning and extra physics is required to explain the delayed radio flare of April 4, 2007 (Yusef-Zadeh et al. 2009) via an adiabatic expansion initiated by the NIR/X-ray flare.

### 3.5. Energetics and the Size of the Emitting Region

The total energy supplied to electrons with $\gamma \gtrsim \gamma_{min} = 10$ in our $p = 2.4$ model that reproduces the bright NIR and X-ray flare from Sgr A* is

$$\Delta E \approx 3 \times 10^{39} \text{ erg}$$

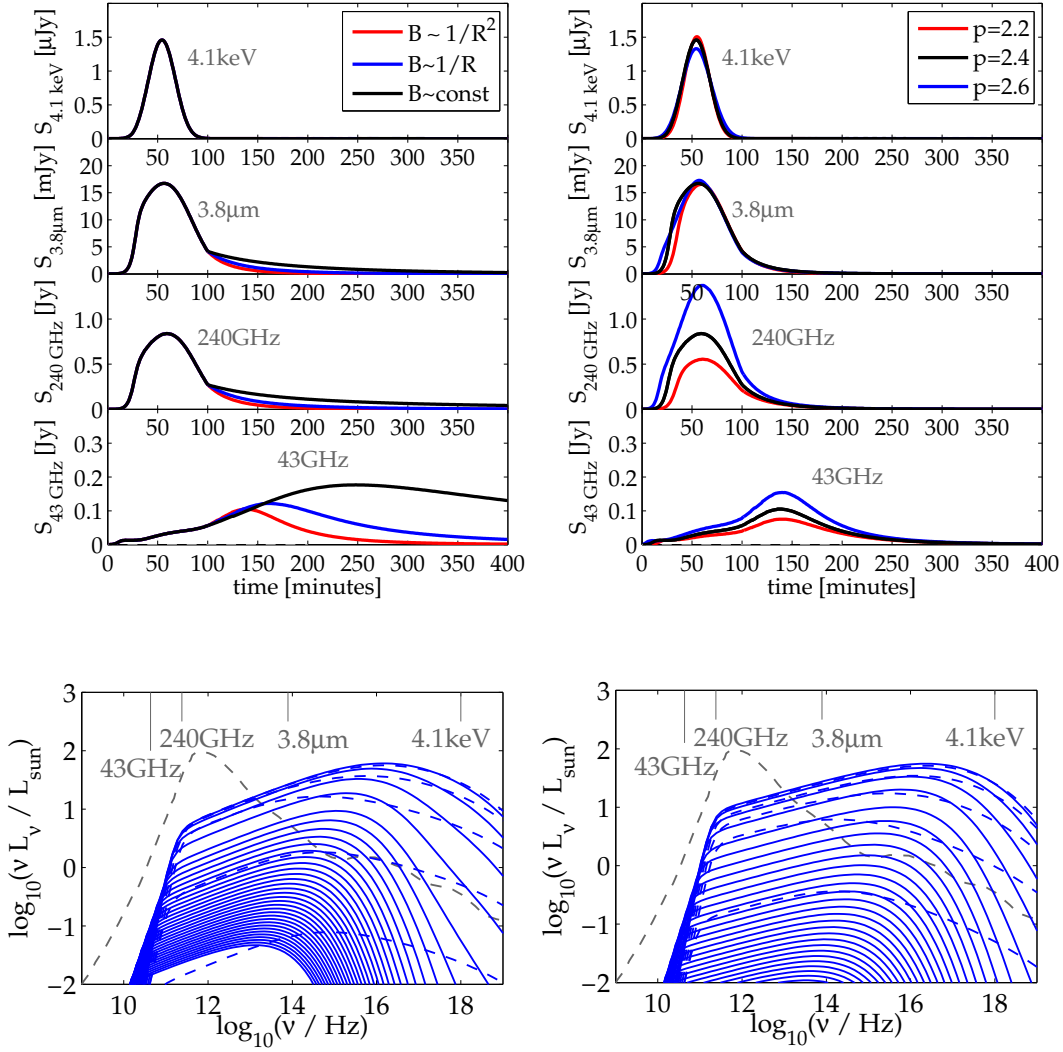

FIG. 8.— Adiabatic expansion models for radio-mm emission. The left panel shows the effect of varying different functions $B(R)$, the right panel the effect of different particle index $p$ (see Figure 7 for the effects of different $R_i$ and $v_{\rm exp}$). For the left panel the model from Section 3.3 with $p = 2.4$, $B(t_{\rm peak}) \simeq 5$G and $t_{\rm esc} = 10000$ min sets the initial conditions for the expansion, which begins at $t = 100$ min. For the right panel $B(t_{\rm peak}) \simeq 11$ G and 2 G for $p = 2.2$ and $p = 2.6$ respectively. When not otherwise listed in the legend the model has $v_{\rm exp} = 0.54 R_i/{\rm hr}$, $R_i = 1 r_g$, and $B \propto 1/R^2$. See caption of Figure 7 for further explanation.

In the same model, the magnetic field decreases from $\sim 40$G to $\sim 5$G (see Figure 5). For the magnetic energy decrease itself to power the flare the decrease must occur in a region with a size

$$R \gtrsim \left(\frac{6\Delta E}{\Delta B^2}\right)^{1/3} \approx 1.5\ r_g, \quad (10)$$

where the equality requires that the magnetic energy is converted into electron energy with 100 % efficiency. A large efficiency may not be unreasonable: in solar flares a large fraction of the released energy appears to go into particle acceleration (Lin et al. 2003). Also note that magnetic dissipation may occur over a volume much bigger than that of the current sheet where particle acceleration happens (see Figs. 1 & 2).

There are independent constraints on the size of the flaring region. For example, the requirement that *hard* synchrotron self-Compton emission should be subdominant in the X-rays (reminding that soft SSC only occurs for extreme magnetic fields and densities; Dodds-Eden et al. 2009) puts a lower limit on the size of the emission region of $R \gtrsim 1 r_g$. More quantitatively, we find the SSC component contributes to 30%, 16% and 9% of the total X-ray flux for $R = 1 r_g, 1.5 r_g$ and $2 r_g$, respectively.[7] An approximate upper limit on the size of the flaring region comes from short timescale variations in the lightcurve of the April 4 event: these constrain 30% of the flux to come from a region with $R \lesssim 1.2 r_g$. In addition, the total number of accelerated electrons required to produce the flare, together with independent estimates of the ambient electron density, favor a size of $R \sim 1.5 - 2 r_g$. Specifically, for $R = 1 r_g$ and $\gamma_{\rm min} = 10$, we require a local density of accelerated particles of $n_e \approx 2 - 8 \times 10^8 {\rm cm}^{-3}$ depending on the escape timescale; this is larger than the ambient density $\sim 10^7$ cm$^{-3}$ estimated from mod-eling the 'quiescent' emission and the observed Faraday rotation (e.g., Yuan et al. 2003). By contrast, for $R = 2r_g$ and/or somewhat higher $\gamma_{\min}$, we find better consistency with the ambient density estimates.

[7] The IC emission produced by upscattering submm photons (assuming $R_{\rm submm} \approx 4 r_g$; Doeleman et al. 2008) contributes only 5-7% of the total X-ray flux for $R = 1 - 2 r_g$, respectively.

Taken together, a flaring region with a size $\simeq 1.5 - 2\ r_g$ is implied by the observed properties of the NIR and X-ray flare, the ambient density constraint, and the energetics of the flare (eq. 10). This is also similar to the size of the region in the MHD simulations in which the magnetic energy decreases dramatically and the plasma is heated (§2).

### 3.6. Effect of the Decreasing Magnetic Field on the Steady State Emission

Our calculations demonstrate that longer wavelength delayed flares from adiabatic expansion of the initially accelerated particles are relatively faint and may be difficult to detect. However, there is another important implication of this model for longer wavelength emission.

Given that the size of the emission region estimated in §3.5 is comparable to the likely size of the sub-mm emitting region, if the magnetic field indeed decreases as we have argued here, the emissivity of the submm-emitting electrons could be significantly reduced (emissivity $\sim B^2$). We thus expect a *reduction* in the quiescent emission at submm wavelengths accompanying the NIR/X-ray flare. Note that this is also consistent with the MHD simulations, in which the magnetic field strength decreases over the entire inner region of the flow (the likely sub-mm emitting region; Fig. 1).

It is intriguing that there is such a dip in the 230/240 GHz emission following the bright NIR/X-ray event on April 4, 2007. This dip can be seen in Figure 10 (discussed below) and lasts for a total of ≈400 minutes. At its lowest the flux reaches $\sim$ 1.7 Jy, significantly below the average mm flux of about 2.8-3 Jy (Zylka et al. 1995; Falcke et al. 1998; Zhao et al. 2003). It is also notable that this is the lowest flux measured for all 230-250 GHz observations (SMA, SMT and IRAM) of the April 2007 campaign (Yusef-Zadeh et al. 2009). After the dip the flux rises again and from 500 minutes reaches fluxes $\sim$3 Jy which are comparable with the average mm flux for Sgr A*. Possibly the radio 'flare' too, which rises around a similar time to the mm lightcurve, is also related to the recovery of the steady state emission, though it less clear that the magnetic field could be reduced over such a fraction of the radio-emitting region as to decrease the radio emission significantly.

If magnetic reconnection – accompanied by a simultaneous decrease in the field strength in the inner accretion flow – is the basis for the flares in Sgr A* in general, then this effect should be present in other flares. Previous work has suggested that submm flares follow NIR/X-ray flares by ∼100 minutes (Marrone et al. 2008; Yusef-Zadeh et al. 2008; Meyer et al. 2008; Eckart et al. 2008). However, our work raises the question of whether these submm variations are really flares at all – or is the rise in submm emission after ∼100 minutes simply the 'recovery' from the decreasing $B$ that initiated the NIR/X-ray flare?

The data from previous observations are reasonably consistent with this interpretation. For example, the X-ray flare with apparent delayed mm/submm emission (1.3mm/850$\mu$m) published in Marrone et al. (2008) and Yusef-Zadeh et al. (2008) (Figs. 3 and 2, respectively) could easily be seen as a dip rather than a delayed flare. It is less clear whether the Ks-band flare at 6:00 UT published in Meyer et al. (2008) followed by a H/K'/L' flare at 8:00 UT simultaneous to observations at 1.3mm (Hornstein et al. 2007; Marrone et al. 2008) works well with a dip interpretation, but the 1.3 mm flux is slightly lower during the 6:00 UT Ks-band flare.[8] Finally, in Eckart et al. (2008) (Fig. 4) there is a bright L'-band flare simultaneous with a pronounced dip in the 870$\mu$m emission observed with LABOCA/APEX. The bright initial peak in the NIR is followed by a sequence of smaller peaks: these subsequent peaks however appear to be accompanied by submm activity.

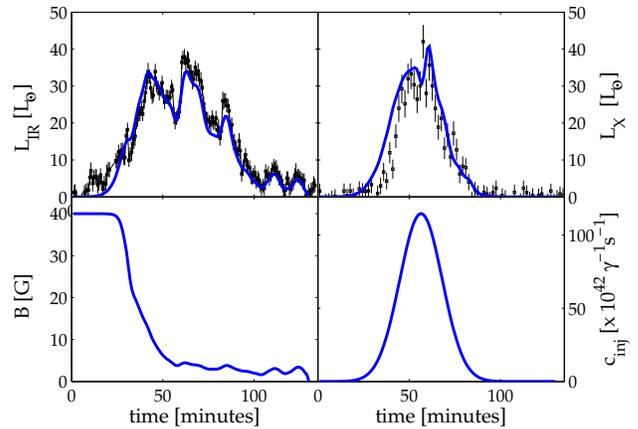

FIG. 9.— The $t_{\rm esc} = 30$ min model of Figure 5, with the addition of small fluctuations in the magnetic field. These can reproduce the substructure seen in the NIR lightcurve, while the X-ray remains relatively unaffected because of the very short cooling time for electrons emitting in the X-ray.

### 3.7. Lightcurve Substructure

Another intriguing feature of the observed April 4 flare is the substructure in the NIR lightcurve that is not present in the X-ray lightcurve. In the synchrotron model, this puzzling property is due to the different response of NIR and X-ray emitting particles to changes in the magnetic field. If there are magnetic field fluctuations, then the infrared emission (below the cooling break) will exhibit significant variations while the X-ray emission will be comparatively smooth because the cooling time for X-ray emitting particles is less than the injection time and so the X-ray luminosity is sensitive largely to the injection rate, not the magnetic field.

To quantify this, Figure 9 shows a model in which we introduce some variation in the magnetic field strength as a function of time (lower left panel); the basic parameters of this model are the same as the $t_{\rm esc} = 30$ min model of Figure 5. The small changes in $B$ in Figure 9 ($\sim 15\%$) can produce the variable NIR emission observed, but they have comparatively little effect on the X-ray emission. Interestingly, the magnetic field changes do introduce some small structure in the X-ray lightcurve as well, at a time when the magnetic field strength is sufficiently low (few G) that the cooling time in the X-ray is comparable to the injection time. This matches a slight feature that is seen in the observed X-ray lightcurve. The same effect may also explain the sharp drops seen in the lightcurves of other bright X-ray flares from Sgr A* (Baganoff et al. 2001; Porquet et al. 2003). These could, of course, equally well be due to changes in particle acceleration. If, however, the result in Figure 9 is correct, it uniquely determines both the magnetic field strength at the peak of the flare (a few G) and the fact that the field strength must have been significantly larger earlier in the flare (on energetic grounds and via the fact that a low field strength throughout the flare is inconsistent with the data; see Fig. 3).

---

[8] this flare is also unusual compared to others in that neither NIR peak had an accompanying X-ray flare, despite simultaneous Chandra observations; and the 1.3mm flux appears to be generally high, ∼4-5 Jy, throughout the observation.

Figure 9 demonstrates that the observed substructure in the lightcurves from Sgr A* can be explained without requiring relativistic effects (that have been suggested previously; e.g., Genzel et al. 2003). Our model also naturally accounts for the different short timescale variability observed in the NIR and X-ray as a consequence of the different synchrotron cooling timescales. The difference in substructure in the NIR and the X-ray lightcurves is not a priori unexpected if the lightcurve variations were due to relativistic effects, where one anticipates both NIR and X-ray emission should undergo similar amounts of beaming. However it is also not clear that the freshly injected electrons (emitting in X-rays) should have the same instantaneous dynamics as the bulk of the slower-cooling electrons (emitting in the NIR). Given that the timescales we are modeling correspond to multiple orbital periods at the last stable orbit, it is likely that some relativistic effects on the lightcurves must be present. Accordingly, because our model does not include any dynamics, we also cannot rule out that relativistic effects are important and modify the emission from what we find here. This will be studied in more detail in future work.

## 4. CONCLUSIONS

We have presented a model for the time-dependent non-thermal emission produced by transiently accelerated electrons in Sgr A*; although these calculations are in principle quite general, we have focused on the origin of the observed NIR and X-ray flares, and the likelihood of coincident or delayed longer wavelength emission. Our model is motivated by the hypothesis that dissipation of magnetic energy powers the flares, as is the case for solar flares and is believed to be the case in other systems like young stellar objects.

We have shown that episodic magnetic reconnection can occur near the last stable circular orbit in (non-relativistic) MHD simulations of accretion onto a central point mass (Figs. 1 & 2). This occurs when oppositely directed magnetic field lines are brought together by rapid inflow near the last stable orbit. The properties and statistics of these reconnection events depend, however, on the magnetic field we initialize in the disk at large radii. Thus a full understanding of whether such reconnection is indeed generic in RIAF models will require a better understanding of the large-scale magnetic field self-consistently generated in the accretion disk.

Motivated by the reconnection hypothesis, we developed a time-dependent, spatially one-zone, model for the acceleration and cooling of relativistic electrons under conditions appropriate to Sgr A*. Our model lacks the time-dependent dynamics and full general relativity of accretion disk simulations (e.g., Dexter, Agol, & Fragile 2009), but treats the electron distribution function in much greater detail. This is, we have argued, critical for understanding the NIR and X-ray emission produced by non-thermal particles.

Our calculations focus on the "cooling break synchrotron" model for the X-ray flares from Sgr A* (Dodds-Eden et al. 2009). In this model, both the NIR and X-ray emission are synchrotron emission. A cooling break between the NIR and X-ray causes the spectrum to steepen by $\Delta \alpha = 0.5$ ($\nu F_\nu \propto \nu^\alpha$); see Figures 3 and 4. This is consistent with the spectral indices of luminous flares from Sgr A*, in particular the very luminous and well-studied flare from April 4, 2007 (Dodds-Eden et al. 2009).

Figure 10 presents the overall picture we have developed for the April 4, 2007 flare. We summarize the findings from our modeling as follows:

### 4.1. Conclusions: NIR and X-ray

Model NIR and X-ray synchrotron lightcurves in which the rise and decay of the emission is governed solely by electron injection and energy loss (e.g., synchrotron cooling, adiabatic expansion, or escape) are either simultaneous and of similar duration, or the NIR lightcurve is delayed relative to, and longer than, the X-ray (the former occurs if the cooling time of NIR-emitting electrons is short compared to the timescale on which relativistic particles are injected, the latter if it is long). Simultaneous lightcurves of different duration - as is observed for luminous flares from Sgr A* – do not occur for fixed plasma parameters during the flare.

The interplay between electron acceleration, synchrotron cooling, and magnetic field evolution during the flare can produce a model that matches both the average SED and NIR/X-ray lightcurves of the luminous flares from Sgr A* (e.g., that of April 4, 2007). In particular, a magnetic field decrease by a factor of $\sim 3 - 10$ accompanying the injection of relativistic particles can explain the observational result that the NIR and X-ray lightcurves are simultaneous, but of different duration (Fig. 5). This is consistent with the hypothesis that magnetic energy dissipation powers the flare in the first place.

Furthermore, small magnetic field fluctuations can reproduce the lightcurve substructure seen in the NIR lightcurve without producing substructures of similar magnitude in the X-ray. This is because the synchrotron cooling time is typically so short in the X-ray that the emission depends primarily on the rate at which electrons are accelerated, and is relatively independent of the magnetic field strength; the same is not true for electrons emitting in the NIR, where the cooling time is longer.

A general decrease in the magnetic field (that is not so smooth, i.e. with fluctuations) can also be responsible for sharp drops observed near the peaks of X-ray flares from Sgr A*, an effect which results from the cooling break reaching X-ray wavelengths. In summary, we find that with the detailed time-dependence of the magnetic field alone (the energy injection may be rather smooth), one can reproduce all the observed time-dependent features of the simultaneous lightcurves.

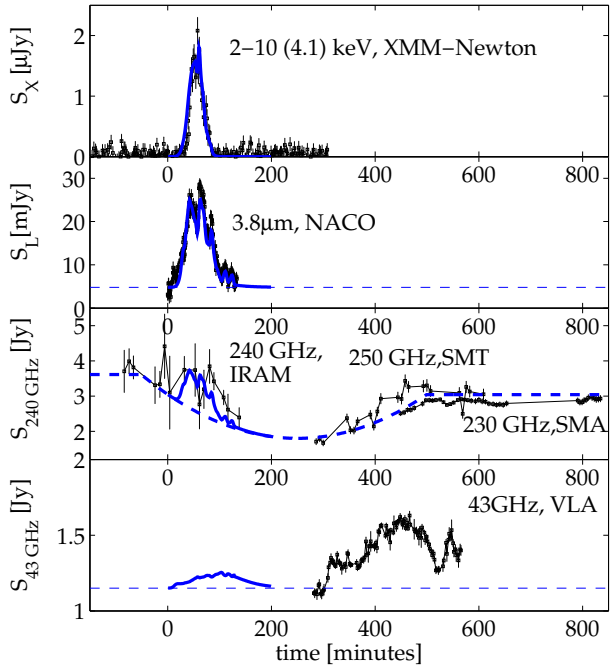

FIG. 10.— Our summary model for the April 4, 2007 flare. The size of the flaring region is $1.5 r_g$. The X-ray and L'-band lightcurves are those shown in Figure 9. We also show the emission at 230 GHz, which is optically thin. In addition to the flare emission we show a schematic lightcurve for the "quiescent" emission (dashed line) which decreases due to the decrease in magnetic field associated with the flare. The steady state emission is re-established after $\sim 400$ minutes. There is a faint ($\sim 0.1$ Jy) radio flare with a delay $\sim 50$ min from the peak of the X-ray/L'-band flare, insufficient to explain the observed radio flare at $\sim 500$ min. The latter may be due to additional particle acceleration in an outflow initiated by the NIR-X-ray flare.

These models predict very little spectral evolution in the NIR and X-ray during the flare, except perhaps some reddening at the very end of the NIR flare, when the flux is $\sim 10\%$ of its peak value (Fig. 5). Marrone et al. (2008) argued that the dominant process governing the rise and decay of the flare emission had to be energy-independent, citing the relative stability of the NIR spectral index with flux (Hornstein et al. 2007), consistent with our finding that there should be very little spectral evolution. However, at the lowest fluxes some authors see a trend towards redder NIR spectra (Gillessen et al. 2006), which in the context of our model could be a sign of the cooling break crossing the NIR bandpass.

### 4.2. Conclusions: Millimeter and radio

We have also studied the emission produced by the adiabatic expansion of plasma during, and after, the NIR and X-ray flare. In our MHD accretion disk simulations, we find that in some cases dissipation of magnetic energy leads to over-pressurized plasma that rapidly expands outwards, analogous to a coronal mass ejection in the sun (Figs. 1 & 2). It is thus possible that the magnetic field decrease required to account for the differences between the NIR and X-ray lightcurves (in the synchrotron model) could be due to outward expansion of the plasma initiated by the flare itself (Fig. 6). In addition to the decrease in magnetic field strength, the magnetic field geometry also changes during the magnetic reconnection event (Fig. 1). This change in magnetic geometry can result in substantial changes in the polarization angle during the NIR flare, as was as observed by Trippe et al. (2007) and Meyer et al. (2006).

In our calculations, we find that the flaring region is unlikely to be self-absorbed at $\sim 240$ GHz. As a result, even though the plasma is expanding outwards at late times, there is no delayed flare at $\sim 240$ GHz. Our models do predict that at $\sim 43$ GHz, with adiabatic expansion there should be a delayed $\sim 0.1 - 0.2$ Jy flare $\sim 100 - 200$ min after the onset of the NIR and X-ray emission (Figures 7 and 8). This is, however, not sufficient to explain the April 4, 2007 radio (43 GHz) flare, which was delayed by 6 hours from the initial NIR/X-ray flare. Furthermore, in our simulations we have not included the additional absorption effect of thermal/non-thermal particles emitting the majority of the emission at submm/radio wavelengths, which could have the effect of suppressing the delayed emission even further, depending on the line of sight to the acceleration region through the accretion flow. These results argue against the radio flares as being produced by outward expulsion of the same relativistic electrons that produced the NIR and X-ray flare. This does not, however, preclude that the radio and NIR/X-ray flares are causally connected. The rise in the radio flux might be related to the disruption in the inner regions of the accretion flow caused by the loss of magnetic energy in the reconnection (as we suggest is the case for the mm emission), or it could be produced by in situ acceleration of particles in an outflow initiated during the NIR/X-ray flare (Liu et al. 2004). Along these lines, it could be reconnection events that are responsible for providing the non-thermal electrons required to produce the relatively flat spectrum observed in the radio for the quiescent state (e.g., Yuan et al. 2003).

Alternatively, it might be that radio flares are unrelated to the high energy particles of the NIR/X-ray flares. For example, it has been shown for the time-dependent jet model of Maitra et al. (2009) that the general spectrum, size measurements and rms variability of Sgr A* (from 7mm to 13 cm), as well as the simultaneous 22 and 43 GHz lightcurves of Yusef-Zadeh et al. (2008) can be explained simultaneously by a jet model with the variations explained by adiabatic expansion of overdensities in the jet. In this case the overdensities would likely arise through variations in the accretion rate, not necessarily linked to the acceleration of particles to high energy that occurs in the NIR/X-ray flares.

Our model makes a strong prediction for the mm emission associated with flares. In a strong magnetic reconnection event, the inner regions of the accretion flow are likely to be disrupted, with the magnetic energy decreasing in a significant portion of the submm-emitting region. We see this explicitly in our MHD accretion disk simulations (Figs. 1 & 2) and the energetics of the luminous flares from Sgr A* support this conclusion (§3.5). After a possible increase in emission due to particles accelerated during the flare, the mm flux should be suppressed by the decrease in the magnetic field in the inner regions of the accretion flow. We argue that there is evidence for such a decrease in the mm observations of the April 4, 2007 flare. The mm flux will recover to its quiescent value when the steady state accretion flow itself readjusts; the timescale for such a recovery is set by the viscous time in the inner parts of the accretion flow which is disrupted. This is somewhat uncertain, but $\sim 3$ hours (Fig. 1). It is likely that the 'dip' and 'delay' in the sub-mm flux will be larger for stronger X-ray/IR flares because the stronger flares correspond to the disruption of a larger part of the quiescent accretion flow.

### 4.3. Final Remarks

Although we believe that the model summarized in Figure 10 is theoretically well-motivated (Fig. 1 & 2) and reproduces the spectral properties and lightcurves of the luminous flares from Sgr A*, it is by no means certain that it is the only explanation. For example, we have assumed throughout this paper that particle acceleration produces a power-law distribution of electrons $\propto \gamma^{-p}$ from $\gamma \sim 10 - 10^6$, with the electron spectral index $p$ independent of time. If however, the injection spectrum varied with time this could in principle alter some of our conclusions.

We have also not considered the possible effects of inverse Compton (IC) cooling on the time-dependent spectrum of electrons. The amount of IC cooling depends on the production rate of inverse Compton scattered photons, which depends on the size of the flaring region. For region sizes $\approx 1.5$-$2 R_S$ (see Section 3.5) synchrotron cooling dominates for the photon densities typical of the flare peak for magnetic fields above $\approx 5 - 7$ G (i.e. while $U_B > U_{ph}$, assuming all scattering is in the Thompson limit). Thus synchrotron cooling will be the dominant effect in the rising phase of the flare while the magnetic field is still high. IC cooling may then start to play some role near the peak of the flare where the photon density is high and the magnetic field has decreased to $\approx 5$ G such that $U_B \approx U_{ph}$. However the magnitude of the effect is not likely to be as large as would be estimated using the approximation from the Thompson limit, since photons scattered from $\gamma_c \sim 10^4$ electrons are already in the Klein-Nishina regime where scattering is less effective for $\nu \gtrsim mc^2/(h\gamma) \approx 10^{16}$ Hz (Rybicki & Lightman 1979). Klein-Nishina effects will then suppress the amount of IC cooling of electrons emitting synchrotron at X-ray wavelengths (depending on the details, the X-ray spectrum may even be fully restored to a synchrotron cooling only regime, see e.g. Nakar et al. 2009). Implementation of IC cooling including the full Klein-Nishina effects requires additional modeling, which we would like to fully explore in future work. Note, however, that (i) the effect of IC cooling produces similar breaks in the spectrum to that of synchrotron cooling so there will be no significant change in the spectrum near the peak of the flare if IC cooling starts to play a role, and (ii) the neglect of IC cooling does not change our main result – the decrease in magnetic field required to explain the simultaneous *rising* phases of the NIR/X-ray lightcurves – since synchrotron cooling dominates in the rising phase of the lightcurve.

Future multi-wavelength observations of flares from Sgr A* will enable us to build statistics and to understand whether the properties of the April 4, 2007 flare are common to Sgr A* flares in general. In the context of synchrotron emission, the fact that the NIR and X-ray lightcurves have different widths and rise times depends on details of the model, such as the escape time or how much the magnetic field decreases during the flare. *A priori* we would thus expect variation in the lightcurve properties from flare to flare. However, one might expect a trend for the peak NIR/X-ray ratio for flares to increase generally for smaller flares, which have a less dramatic magnetic field decrease. For smaller flares, it might then be possible that inverse Compton emission with a harder spectral index dominates the X-ray emission instead.

The relative spectra in the NIR and X-ray are also critical for constraining the theoretical models: it is primarily the combination of the hard NIR spectrum ($\nu L_\nu \propto \nu^{0.4}$) and the soft X-ray spectrum ($\nu L_\nu \propto \nu^{-0.3}$) that rules out IC emission as the origin of the luminous X-ray flares, favoring synchrotron emission instead (Dodds-Eden et al. 2009). The hard NIR spectrum is also what requires efficient acceleration of non-thermal electrons with $p \simeq 2 - 2.4$ near $\gamma \sim 10^3$ ($n(\gamma) \propto \gamma^{-p}$), rather than simply a modest extension of the mm-emitting thermal distribution function (which would predict a relatively red NIR spectrum; Yuan et al. 2003). It is thus critical to understand the spectrum of the NIR emission and whether it depends on flux. At longer wavelengths, it is particularly important to understand the magnitude of the submm flux during and just after NIR/X-ray flares, as compared to times of no NIR or X-ray activity.

In the long term, better understanding the flares from Sgr A* will hopefully enable us to use such time dependent emission as a probe of accretion and outflow physics, and potentially strong gravity. Moreover, understanding the flaring emission in addition to the quasi-steady emission will further refine what physics must be included in time-dependent general relativistic MHD simulations (i.e., as concerns the production of non-thermal particles) in order to explain and predict the emission from Sgr A*.

Support for this work was provided by NASA through Chandra Postdoctoral Fellowship grant number PF8-90054 to PS, awarded by the Chandra X-ray Center, which is operated by the Smithsonian Astrophysical Observatory for NASA under contract NAS8-03060. EQ was supported in part by the David and Lucile Packard Foundation, NSF-DOE Grant PHY-0812811, and NSF ATM-0752503.